\documentclass[runningheads]{llncs}

\usepackage[T1]{fontenc}
\usepackage{graphicx,verbatim}
\usepackage{hyperref}
\usepackage{color}

\usepackage{cite}
\usepackage{float}
\usepackage{tikz}
\usepackage{appendix}
\usepackage{authblk}

\begin{document}
\title{Towards Diagnostic Quality Flat-Panel Detector CT Imaging Using Diffusion Models}
\titlerunning{Towards Diagnostic Quality FDCT Using Diffusion Models}

\author{
Hélène Corbaz\inst{1}\orcidID{0009-0006-9716-5136} helene.corbaz@unibas.ch \and
Anh Nguyen\inst{2} \and
Victor Schulze-Zachau\inst{2}\orcidID{0000-0003-0945-8379}\and
Paul Friedrich\inst{1}\orcidID{0000-0003-3653-5624}\and
Alicia Durrer\inst{1}\orcidID{0009-0007-8970-909X}\and
Florentin Bieder\inst{1}\orcidID{0000-0001-9558-0623}\and
Philippe C. Cattin\inst{1}\orcidID{0000-0001-8785-2713}\and
Marios N. Psychogios\inst{2}\orcidID{0000-0002-0016-414X} 
}
\authorrunning{Hélène Corbaz et al.}
\institute{Department of Biomedical Engineering, University of Basel \and Department of Neuroradiology, University Hospital of Basel  }

\maketitle             
\begin{abstract}
Patients undergoing a mechanical thrombectomy procedure usually have a multi-detector CT (MDCT) scan before and after the intervention. The image quality of the flat panel detector CT (FDCT) present in the intervention room is generally much lower than that of a MDCT due to significant artifacts. However, using only FDCT images could improve patient management as the patient would not need to be moved to the MDCT room. Several studies have evaluated the potential use of FDCT imaging alone and the time that could be saved by acquiring the images before and/or after the intervention only with the FDCT. This study proposes using a denoising diffusion probabilistic model (DDPM) to improve the image quality of FDCT scans, making them comparable to MDCT scans. Clinicans evaluated FDCT, MDCT, and our model's predictions for diagnostic purposes using a questionnaire. The DDPM eliminated most artifacts and improved anatomical visibility without reducing bleeding detection, provided that the input FDCT image quality is not too low. Our code can be found on \href{https://github.com/HeleneCorbaz/SWITCH-2025-FDCT-DDPM}{github}.


\end{abstract}

\section{Introduction}
\begin{figure}[ht]
    \begin{center}
    \includegraphics[width=\textwidth]{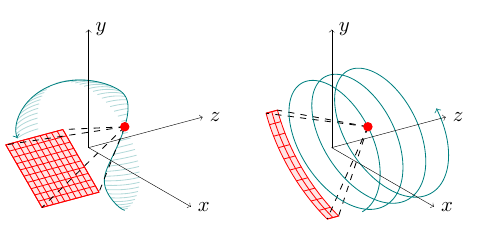}
    \end{center}
    \caption{The difference in acquisition geometry for FDCT (\textit{left}) and MDCT (\textit{right}). The MDCT consists of several arc-shaped detectors. Its acquisition trajectory is helicoidal with multiple rotations. In contrast, the FDCT has one flat detector and can be acquired with a C-arm in the intervention room. Its acquisition is sinusoidal with only one rotation. Although FDCT images can have higher resolution than MDCT images, they are more sensitive to motion and beam hardening artifacts and can present ring artifacts.}
    \label{fig:tech}
\end{figure}

\begin{figure}[ht]
    \begin{center}
    \includegraphics[width=\textwidth]{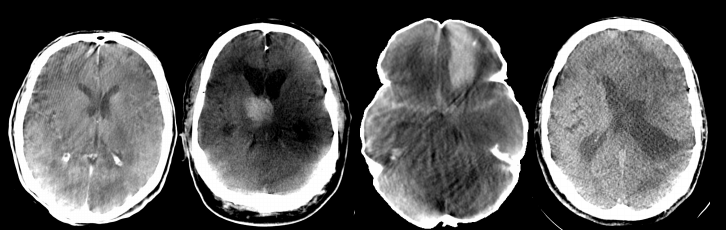}
    \end{center}
    \caption{Typical artifacts found in FDCT images. From \textit{left} to \textit{right}: beam hardening, contrast inhomogeneity, patient's motion, ring artifact. Due to major technical differences, these artifacts are more prevalent in FDCT images than in MDCT images.}
    \label{fig:art}
\end{figure}

In the field of interventional radiology, cone beam computed tomography (CBCT) is also referred to as flat-panel detector computed tomography (FDCT). FDCT has emerged as a valuable imaging technique due to its ability to generate high-resolution three-dimensional images and its capability of being integrated in the C-arm of a fluoroscopy unit\cite{Gupta2008-gl, Kalender2007-hl}. In interventional neuroradiology, FDCT is often used during or immediately after procedures. This enables the rapid detection of intracranial hemorrhages and the verification of intravascular device placement, such as stents. FDCT is particularly valuable following mechanical thrombectomy, a procedure performed to recanalize occluded arteries in patients with acute ischemic stroke\cite{Turc2023-ig, Jadhav2021-eu}.
However, a significant limitation of FDCT compared to multidetector CT (MDCT)  (see Figure \ref{fig:tech}) is its higher susceptibility to artifacts, including beam-hardening, undersampling-induced patterns, motion artifacts, and cone-beam geometry artifacts (see Figure \ref{fig:art}). Additionally, detector defects are more prevalent with FDCT than with MDCT\cite{wei_reduction_2024, cancelliere_butterfly_2023, kachelries_advanced_2000}. These artifacts can obscure essential anatomical details and potentially result in diagnostic errors\cite{barrett_artifacts_2004}.\\ 
In the context of interventional neuroradiology, retrospective studies have demonstrated the feasibility of using FDCT to identify early ischemic lesions and intracranial hemorrhages with a level of diagnostic accuracy comparable to that of MDCT\cite{Leyhe2017-gi, Maier2018-ng}. However, FDCT remains inferior to MDCT in terms of image quality, especially with regard to gray-white matter differentiation and visualization of infratentorial structures, primarily due to pronounced beam-hardening artifacts.\\
To address these limitations, the non-inferiority SPINNERS trial\cite{trial} is evaluating whether FDCT alone can distinguish ischemic from hemorrhagic stroke with accuracy comparable to MDCT. If FDCT proves to be non-inferior, it could be adopted into streamlined ``one-stop'' or direct-to-angiography pathways. This would bypass conventional MDCT in the emergency department and enable immediate transfer to the angiography suite for imaging and, when indicated, endovascular therapy. Since door-to-treatment time is strongly associated with clinical outcomes, as demonstrated by a post hoc meta-analysis of five trials\cite{Saver2016-sz}, this strategy could reduce treatment initiation time by several minutes and improve functional outcomes\cite{Brehm2019-mz, Psychogios2019-wd, Requena2020-ca, Mendez2018-sa}.

\subsubsection{Related Work}
Several studies have demonstrated the potential of generative models to improve the quality of FDCT images. For example, Chen et al.\cite{chen} used a U-Net-based architecture, and Zhang et al.\cite{zhang} used a GAN to enhance radiation therapy image quality. In the SynthRad2023 challenge\cite{synthrad2023}, different methods based on CNNs, transformers, GANs, and diffusion models were employed to perform image-to-image translation tasks: MRI to MDCT translation for the first task and FDCT to MDCT translation for the second task. In the first task, the first and the second place used a transformer and a CNN respectively, In the second task, the first place used a CNN and the second place used a transformer. Diffusion models were in the middle of the ranking for both tasks. GANs and other CNNs held the last positions for both tasks.
However, given the excellent performance of DDPMs in other image-to-image translation tasks \cite{saharia_palette_2022, saharia_image_2023,friedrich_cwdm_2024, friedrich_deep_2025, durrer_diffusion_2023}, we aimed to further investigate this approach.

\subsubsection{Contribution}
To enhance the diagnostic potential of FDCT imaging, we propose a 2D diffusion-based approach that maintains the high resolution ($512\times 512$) of the FDCT images while eliminating artifacts and improving contrast. In addition to commonly used quantitative evaluation metrics, clinicians rated the image quality and diagnostic potential compared to original FDCT and MDCT images. The DDPM significantly reduced artifacts and enhanced anatomical visibility while still allowing effective bleeding detection, given that the quality of the input FDCT images is sufficiently high.

\section{Methods and Experimental Details}
Similarly to Durrer et al. \cite{durrer_diffusion_2023}, we performed image-to-image translation using MDCT images as the target and FDCT images as the input. We adapted a DDPM implementation by \emph{MONAI Generative}\cite{monai_generative}. DDPMs have two main components. First, noise is added to samples of a data distribution (in this case, images) over a certain number of time steps until we reach a standard normal distribution. This is known as the \emph{forward} diffusion process. The model's aim is to learn the \emph{reverse} process and generate the target image by removing the noise step-by-step. To perform the image-to-image translation task, we conditioned the model on FDCT images to reconstruct the paired MDCT image\cite{Ho_ddpm, wolleb_diffusion_2022}. For comparison, we tested pix2pix\cite{pix2pix2017}, a GAN-based method for paired images.

\subsubsection{SPINNERS Dataset} We collected FDCT and MDCT image data from 194 patients in an ongoing, international, multicenter clinical trial. The inclusion criteria and the patient characteristics can be found on the trial page\cite{trial}. Additionally, we excluded 19 cases in which the patient refused consent for further research (7 cases), in which the FDCT and/or MDCT images were missing (6 cases), or in which image registration (2 cases) or skull stripping failed (4 cases), cf. \emph{Preprocessing} below.  We used 125 cases for model training and 50 cases for testing and clinical evaluation.

\subsubsection{SynthRad2023 Dataset}
For comparison, we also used SynthRad2023 challenge images from Task 2, which is image-to-image translation from FDCT to MDCT. SynthRad2023 is a publicly available dataset containing 180 paired head FDCT and MDCT images acquired on radiation therapy machines. Since we do not have access to the challenge's test dataset, we divided the available training set to create training and test images. We used 158 images for training and 22 for testing. Therefore, our results can not be directly compared to those detailed in the challenge report\cite{synthrad_report}.

\subsubsection{Preprocessing}
For the SPINNERS dataset, we coregistered the FDCT and MDCT images using the \emph{Advanced Normalization Tools} (ANTs) library\cite{ants}. Then, we skull-stripped the images using SynthStrip\cite{SynthStrip}. We removed empty slices from both volumes of the FDCT and MDCT images. Finally, we windowed the images at a level of 50 with a width of 100, and then normalized them to the interval [0, 1]. The original image size was between $512 \times 512$ and $1024 \times 1024$. We resized all images to a resolution of $512 \times 512$. For the SynthRad2023 dataset, the original image size was between $270 \times 512$ and $800 \times 512$. We resized all images to $256 \times 256$. The images were windowed at level 50 with a width of 100, and were then normalized to the interval [0, 1].

\subsubsection{Implementation Details}
For the DDPM architecture, we tried to create the largest possible model given the hardware constraints. We used nine resolutions, with (64, 128, 256, 256, 256, 256, 256, 256, 256) channels. Attention was only applied on the lowest resolution. We employed a 64-head transformer. We used 1000 sampling steps and the Adam optimizer, with a learning rate of $2.5 \cdot 10^{-5}$ and a batch size of 12. The DDPM was trained for 642 epochs for the SPINNERS images as the performances were not further improving and for 1000 epochs for the SynthRad images on an NVIDIA A100 40GB GPU. The model's training required 39\,GB for the SPINNERS dataset and 32\, GB for the SynthRad dataset. We employed Python 3.10.15 and PyTorch 2.4.0 as software framework.

\subsubsection{Evaluation}
We evaluated the performance of the model using the mean square error (MSE), the peak signal-to-noise ratio (PSNR), and the structural similarity index measure (SSIM) of the reconstructed images, that we stacked to reconstruct 3D volumes. We converted back the pixels values to Hounsfield Units. Additionally, we conducted a clinical evaluation of ten of the SPINNERS cases in which two neuroradiologists (one junior and one senior) evaluated the FDCT, MDCT, and model prediction (see Figure~\ref{fig:results}) blindly using 31 questions similar to those in Leyhe et al.\cite{Leyhe2017-gi}. We asked the neuroradiologists to evaluate the presence of artifacts (see Figure~ \ref{fig:art}), the visibility of bleeding, and the visibility of anatomy. When possible, we also asked the neuroradiologists to assess the \emph{Alberta Stroke Program Early CT Score} (ASPECT)\cite{aspect}, a tool used to evaluate the extent of early ischemic changes in patients with ischemic strokes. This second clinical analysis is crucial because we do not expect the commonly used metrics PSNR and SSIM to reflect the clinical value of the model's predictions.

\section{Results and Discussion}

\subsection{Quantitative Performances}

\begin{table}
\centering
\caption{Results on the test sets. We report mean values $\pm$ standard deviation.}\label{tab1}
\begin{tabular}{|l|l|l|l|}
\hline
Experiment &  MSE [$HU^2$] $(\downarrow)$ & SSIM $(\uparrow)$& PNSR $(\uparrow)$\\
\hline
DDPM with SPINNERS images &  \hphantom{0}78.22 $\pm$ 0.78  & 0.8568 $\pm$ 0.0007  & 21.45 $\pm$ 0.04\\
DDPM with SynthRad2023 images & 122.27 $\pm$ 3.09 & 0.8527 $\pm$ 0.0029 & 19.58 $\pm$ 0.09\\
Pix2Pix with SPINNERS images & 513.61 $\pm$ 1.81 & 0.7998 $\pm$ 0.0007 & 12.96 $\pm$ 0.01\\
\hline
\end{tabular}
\end{table}

\begin{figure}[ht]
    \begin{center}
        \includegraphics[width=\textwidth, trim=1 0 1 0, clip]{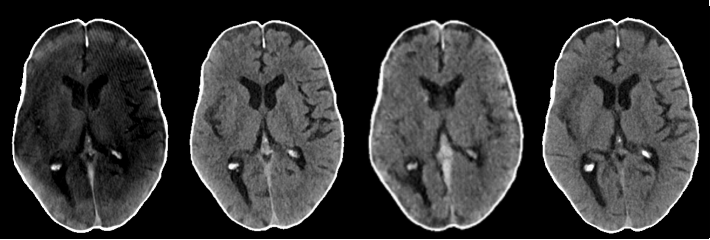}
    \end{center}
    \caption{Example of the results of the models on an image from the test set. From \textit{left} to \textit {right}: Original FDCT image (input), DDPM's prediction, pix2pix's prediction, MDCT image (target). 
In general, pix2pix's predictions show significant anatomical changes, whereas the DDPM's predictions better preserve the anatomy overall.}
    \label{fig:results}
\end{figure}

We trained our models in 2D due to hardware limitations. Some minor inhomogeneities between slices were noticed. We achieved an SSIM of 0.86 and a PSNR of 21.45 on our test set (see Table~\ref{tab1}). Due to the high degree of heterogeneity of the data collected for the SPINNERS trial, we expected the quality of the DDPM's predictions to be heterogeneous, which would result in a high standard deviation (STD). However, the STD across all metrics was low, and the DDPM appears robust to these differences. Pix2pix produced lower-quality images and did not meet the clinicians' expectations regarding general image quality (see Figure~\ref{fig:results}). In some cases, pix2pix's predictions present major changes in the anatomy. We also trained the DDPM on SynthRad2023 challenge images. Our results differed from those of the proposed DDPM during the challenge (SSIM of 0.85 vs 0.88)\cite{synthrad_report}. These differences can be explained by the different training and test sets, as well as different preprocessing steps (skull-stripping and windowing to brain soft tissues). However, these metrics do not reflect the diagnostic value of the images, as they do not account for clinical relevance, the ability to detect specific pathologies, or the overall interpretability by medical professionals. Thus, we additionally performed a qualitative analysis on a subset of ten cases from the SPINNERS test set, selected to contain images from different centers. We decided not to conduct any further qualitative analysis for pix2pix. 

\subsection{Qualitative Performances}

\subsubsection{Artifact Reduction}
All ten FDCT images contained artifacts of varying intensity. However, the presence of these artifacts did not always impact the diagnostic quality of the images. MDCT images exhibited milder artifacts, primarily beam hardening, and light contrast inhomogeneity. This can be explained by MDCT's advanced artifact reduction tools, such as metal artifact reduction, which are used in post-processing. The DDPM predictions showed contrast inhomogeneity artifacts and mild beam hardening. In general, the model's predictions presented fewer artifacts than the target MDCT images themselves. All ring artifacts were removed, and beam hardening was improved. However, the model struggled to correct for contrast inhomogeneity; the model's predictions still contained stronger contrast inhomogeneity than the MDCT images. Overall, the presence of beam hardening and contrast inhomogeneity in the model's predictions seems to depend greatly on their intensity in the FDCT images. Only one case presented a motion artifact. Since this case had strong artifacts overall, the model's prediction did not produce good results when viewing the anatomy, even though it reduced the artifacts. Thus, we cannot draw any conclusions about the model's efficiency in reducing motion artifacts. Another observation is that our model reduces the noise level in the images. However, noise itself is not considered an artifact.  Overall, the DDPM effectively reduced the artifacts present in the FDCT images, especially ring and beam hardening artifacts. Fewer artifacts can help clinicians focus more easily on the diagnostic aspects of an image.

\subsubsection{Anatomy}
Overall, we observed an improvement in the visibility of the brain's anatomy in the DDPM's predictions compared to the FDCT images. However, the predictions still do not match the quality of MDCT images, especially for the differentiation between grey and white matter. We also observed some gyri with unexpected shapes in the DDPM's predictions. We assume that this is the case when there is no information about the anatomy in the FDCT image due to strong artifacts, especially strong beam hardening, leading the DDPM to generate inaccurate structures. We also asked the clinicians to search for hyperdense vessel signs\cite{Gacs1983} in the images. In two cases, a hyperdense vessel sign was visible in the MDCT images and the DDPM's prediction, but not in the FDCT images. In two cases, a hyperdense vessel sign was visible in the MDCT images, but the two raters disagreed on whether it was visible in the other images. In one case, it was only visible in the DDPM's prediction, which was probably a hallucination of the DDPM.

\subsubsection{Bleeding}
There was an agreement on the presence of bleeding in most of the cases. In three cases, the raters perceived the quality of the DDPM's prediction to rule out any type of bleeding differently. For one case, the raters did not agree on the presence or absence of bleeding in the FDCT, but agreed that bleeding was absent in the MDCT and the DDPM's prediction. For one case, the rater evaluated the bleeding to be larger in the DDPM's predictions as for the two other modalities. For one case, one rater rated the bleeding in the MDCT to be larger as in the two other modalities. In one case, the bleeding was visible in the FDCT and MDCT images but not in the DDPM's prediction. This discrepancy is likely due to the poor quality of the FDCT image and might have led the DDPM to interpret it as an artifact. Overall, there is an agreement on the location of the bleeding between the three modalities. Nevertheless, more cases are needed to estimate how often the DDPM makes such mistakes and to evaluate whether it only occurs when the FDCT image's quality is very low. FDCT images appear to have good diagnostic potential for ruling out bleeding after thrombectomy. It seems like DDPM predictions tend to preserve bleeding if FDCT artifacts are not too strong.

\subsubsection{ASPECT}
In five cases, both raters found no signs of early ischemia in the three modalities. In one case, one rater evaluated an average ASPECT score of 9 for MDCT and 10 for the other two modalities. In another case, the average ASPECT score was 8.5 for the DDPM prediction and 10 for the other modalities. This is likely an example of a DDPM hallucination. In another case, the average ASPECT score was 8 for the MDCT images and 9 for the other two modalities. One rater gave the MDCT images a score of 10, while the other gave them a score of 6. Both raters agreed on a score of 6 for the MDCT images and a score of 10 for the model's prediction. Finally, one rater found a score of 3 and the other found a score of 8 for the MDCT images. They agreed on a score of 10 for the other two modalities. Of the ten cases, two presented obvious early ischemic signs on the MDCT images, including potential large infarcts. However, the model returned images without lesions.The average difference between the DDPM's prediction and the MDCT images for the other eight cases was less than two. It would be interesting to see how the model performs with larger ischemic lesions, as only one example was present in this subset. 
\\
We explain the differences in scoring between the MDCT images and the other two modalities due to the time gap of up to four hours between acquiring the FDCT image and the MDCT images. This time gap allows for the occurrence of physiological and pathological processes. By the time MDCT images are acquired, ischemic processes are likely in a more advanced stage, making them more visible. Regarding the DDPM's predictions, however, the visibility of the ischemic lesion did not improve. \\
Lesions visible in MDCT images but not in FDCT images were not necessarily more visible in DDPM predictions. This is likely due to the lack of FDCT information. If a lesion is not visible in the FDCT due to artifacts or because it is too early to detect, the DDPM cannot display it in its predictions. This is not a limitation of our model, but rather a limitation of the FDCT input. On the other hand, the DDPM attempts to correct for contrast inhomogeneity. However, it may not fully capture the difference between a contrast change due to an artifact and a contrast change due to an ischemic lesion, as these changes can be subtle. Having more training examples that better represent these changes would provide more accurate predictions.\\
 
\section{Conclusion}
The quality of FDCT images can be improved using DDPMs. With our model, we could remove most of the artifacts present in the images and even achieved fewer artifacts than MDCT in our test set. 
As long as the input FDCT images are of sufficient quality, the DDPM improves the visibility of the anatomy while still enabling the detection of bleeding. 
However, the DDPM can potentially hallucinate, particularly if severe artifacts are present in the original FDCT images.
For future work, we plan to include estimates of uncertainty in the predictions of the network to account for potential hallucinations of the model. A classifier could be trained to detect FDCT images with a too low quality and higher risk of hallucinations.
We planned to realize a more detailed qualitative analysis of the 50 images of the SPINNERS dataset's validation set in collaboration with clinicians to clarify our remaining incertitude, especially for the detection of early ischemic signs for large infarct and the impact of the quality of the input FDCT to better assess to diagnostic potential of our method and the clinical implication. An evaluation of the model on an external dataset could be performed.

\subsubsection{Acknowledgments} We would like to express our sincere gratitude to the investigators of the SPINNERS clinical trial.

\subsubsection{Disclosure of Interests}
Victor Schulze-Zachau is the recipient of research grants from Basel University, Basel, Switzerland; Bangerter-Rhyner-Foundation, Basel, Switzerland; and Freiwillige Akademische Gesellschaft Basel, Basel, Switzerland. Marios Nikos Psychogios is the recipient of research grants from the Swiss National Science Foundation (SNF) for the DISTAL trial (33IC30\_198783), ICARUS (32003B\_220118) and TECNO trial (32003B\_204977); Bangerter-Rhyner Stiftung for the DISTAL trial ; Unrestricted Grants for the DISTAL trial from Stryker Neurovascular Inc., Medtronic Inc., Phenox GmbH, Penumbra Inc. and Rapid Medical Inc. ; Unrestricted Grant for the ICARUS trial from Acandis GmbH; Sponsor-PI SPINNERS trial (Funded by a Siemens Healthineers AG Grant), Research agreement with Siemens Healthineers AG, Local PI for the ASSIST, EXCELLENT, TENSION, COATING, SURF and ESCAPE-NEXT trials. Speaker fees: Stryker Neurovascular Inc., Medtronic Inc., Penumbra Inc., Acandis GmbH, Phenox GmbH, Siemens Healthineers AG.

\newpage
%
%
%
%
\bibliographystyle{splncs04}
\bibliography{bib.bib}

\newpage 

\end{document}